\documentclass{article}

\usepackage{arxiv}

\usepackage[utf8]{inputenc} 
\usepackage[T1]{fontenc}    
\usepackage{hyperref}       
\usepackage{url}            
\usepackage{booktabs}       
\usepackage{amsfonts}       
\usepackage{nicefrac}       
\usepackage{microtype}      
\usepackage{lipsum}		
\usepackage{graphicx}
\usepackage{natbib}
\usepackage{doi}

\title{ExpTrialMng: A Universal Experiment Trial Manager for AR/VR/MR Experiments based on Unity}

\author{ {Jinwook Kim}\\
	Graduate School of Culture Technology\\
	Korea Advanced Institute of Science and Technology\\
	South Korea, Daejeon \\
	\texttt{jinwook.kim31@kaist.ac.kr} \\
	\And
	{Yee Joon Kim}\\
	Institute for Basic Science\\
	South Korea, Daejeon \\
	\texttt{joon@ibs.re.kr} \\
	\And
	{Jeongmi Lee}\\
	Graduate School of Culture Technology\\
	Korea Advanced Institute of Science and Technology\\
	South Korea, Daejeon \\
	\texttt{jeongmi@kaist.ac.kr} \\
}

\renewcommand{\shorttitle}{ExpTrialMng}

\hypersetup{
    pdftitle={ExpTrialMng: A Universal Experiment Trial Manager for AR/VR/MR Experiments based on Unity},
    pdfsubject={cs.HC},
    pdfauthor={J. Kim, Y.J. Kim, J. Lee},
    pdfkeywords={cognitive psychology experiment, Trial Management, Unity, 3D Environment},
}

\begin{document}
\maketitle

\begin{abstract}
	Based on the improvement of recent virtual and augmented reality (VR and AR) Head Mounted Display (HMD), there have been attempts to adopt VR and AR in various fields. Since VR and AR could provide more immersive experimental environments and stimuli than 2D settings in a cost-efficient way, psychological and cognitive researchers are particularly interested in using these platforms. However, there is still an entry barrier for researchers who are not familiar with Unity programming, and current VR/AR HMDs could also cause unexpected errors during the experiment. Therefore, we developed a Unity library that can be adopted in various experiments universally and assist researchers in developing their own. Our library provides functions related to trial assignment and results saving. That way, researchers can easily implement the essential functions of their psychological experiments. We also made a function that enables proceeding with the experiment from a specific trial point to handle unexpected errors caused by HMD tracking loss issues during the experiment. We expect our library could invite researchers from various disciplines and help them acquire valuable insights in VR/AR environments.
\end{abstract}


\section{Introduction}
With the development of immersive devices such as virtual reality and augmented reality, research is being attempted from a psychological and design perspective rather than a technical point of view.~\cite{kim2022effect, peng2020walkingvibe, tregillus2016vr}. The virtual environment and interactions are mostly developed with 3D programming tools, such as Unity or Unreal. Compared to PsychoPy, which is a python library that provides GUI programming for developing psychological experiment~\cite{peirce2019psychopy2}, these tools require high proficiency in programming skills to develop the whole experiment structure. These features could act as an entry barrier for researchers who conduct user experience or experimental studies based on the design or cognitive science perspective, who are mostly not familiar with 3D programming tools.

In order to assist these researchers, various frameworks and tool-kits provide source codes on Github that could reduce the effort of implementing experiments or stimuli in Unity~\cite{brookes2020studying, zenner2021hart}. However, those frameworks include heavy functions, such as server connection, which are unnecessary for some experimental circumstances. Sometimes this makes it harder to implement experiments from zero bases. Excluding unused functions could cost more time and effort and lead to serious compatibility issues later. In addition, those frameworks do not provide functions that resume the experiment from a specific point. For virtual reality (VR) experiments, various components could cause errors (i.e., tracking, power and connection issues, etc.) in the middle of the experiment, and in that case, it needs to be restarted at that specific point of error, instead of starting from the beginning again.

\begin{figure}
  \includegraphics[width=\textwidth]{figures/expMngFlow.png}
  \caption{The operation flow of ExpTrialMng. Based on the pre-generated CSV file that includes all trial information, it delivers each trial setting to the experiment manager. After a trial is completed, it saves the data to an additional CSV file and loads the following trial information. In addition, it starts the trial from the point where the results are not filled.}
  \label{fig:teaser}
\end{figure}

Therefore, we developed a Unity library\footnote{\url{https://github.com/jinwook31/Unity-Experiment-Trial-Manager}} that provides essential functions for most Human-Computer Interaction (HCI) studies. It could be adapted universally to VR/AR-based experiments easily. Our Library includes a restart function from a specific trial number where an error stopped an experiment, and the trial randomization process is separated by reading the pre-generated CSV file. Also, we implemented a function that saves results from each trial into a single CSV file. We expect that this library could help researchers focus on their experiment and stimuli design and ease the burden on those who are not familiar with Unity programming. Eventually, it could invite researchers from various disciplines into VR/AR and achieve a broader perspective. We propose that our library makes the following contributions:

\begin{itemize}
  \item It lowers the barriers to conducting HCI experiments within VR/AR environments for researchers who are unfamiliar with Unity programming.
  
  \item It provides a general-purpose package that can be applied to diverse experiment settings with minimized lines of code.
  
  \item It minimizes data loss due to unexpected errors and allows easy resumption by modifying the pre-generated trial data file based on the point of interruption.
\end{itemize}

\section{Stimuli Presentation Library}
Most psychological and design user experiments commonly require functions that present stimuli in a randomized order and save the data into a specific format for analysis~\cite{kim2022effect, wang2022movement, bae2018dissociable, gehrke2019detecting}. Therefore, we focused on these functions and tried to make them universal in various experiment settings. Also, we minimized the effort of adapting it to user experiments. Our library's requirements and operation flow are as follows (see Figure~\ref{fig:teaser}).

\begin{figure}
  \includegraphics[width=1\textwidth]{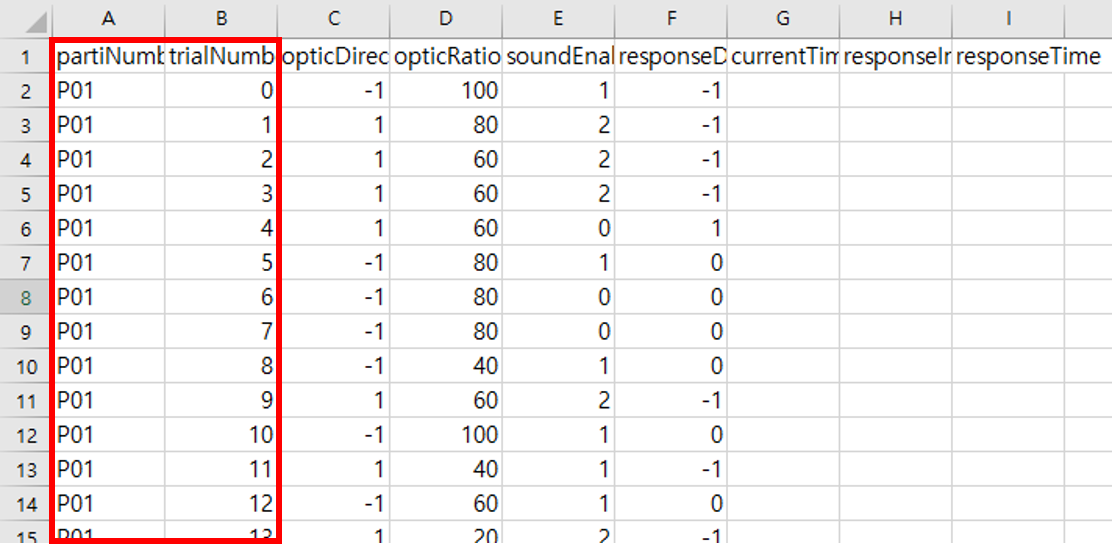}
  \caption{An example of a pre-generated CSV file that includes trial information. The first and second columns must be partiNumber and trialNumber for the library to operate correctly. Column A to F indicates the inputs regarding stimuli settings for each trial, and column G to I is defined as output columns to save the experiment results.}
  \label{fig:preCSV}
\end{figure}

\subsection{Pre-generated Trial Data}
Before starting the experiment, the researcher needs to generate a CSV file containing stimuli information for each trial and place it in the Asset folder (see Figure~\ref{fig:preCSV}). We separated this part due to the various methods for randomization. Therefore, researchers must generate the appropriate CSV file for their experiment. The CSV must include participant number (partiNumber) and trial number (trialNumber) in the first and second header index in order to initiate the experiment. Also, the slots for output need to be empty. It will be filled when the trial is done, or else it will be skipped when the experiment is restarted.

\subsection{Experiment Manager}
\begin{figure}
  \includegraphics[width=0.7\textwidth]{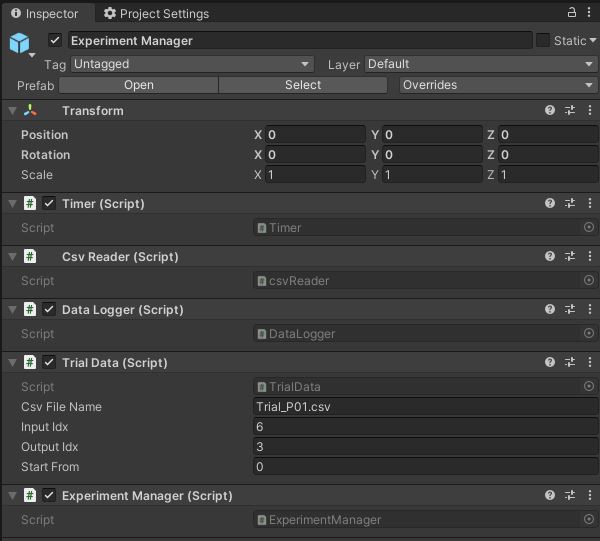}
  \caption{The connected scripts in 'Experiment Manager' Unity prefab. Users need to configure the appropriate pre-generated CSV file name, the number of input and output features in the CSV header, and startFrom index indicating the trial position where the experiment will start in TrialData tab.}
  \label{fig:expMng}
\end{figure}

As Figure~\ref{fig:expMng}, researchers need to import the Experiment Manager prefab to their Unity project and set the appropriate parameters in the TrialData tab. After setting the parameters, the researcher needs to link the code that activates and manages stimuli (e.g., visual, audio, haptic, etc.) with the trial information in the 'ExperimentManager.cs' script. We wrote an example code that could get the trial settings from a pre-generated CSV file and iterate the trials until it is ended. Researchers do not need to look into the code in detail and easily adapt these functions by using a few lines of code. When the experiment is done, they can find the results recorded in a CSV file format in the Asset folder.

\section{Discussion \& Conclusion}
In ExpTrialMng, we implemented a Unity library that includes only the core functions for psychological and design user test experiments and enables researchers adopt it to their experiment easily and universally. The library delivers the trial information from a pre-generated CSV file and saves the experiment results into an additional file. Also, it includes a function that enables the experiment to restart from a specific trial in case of unexpected errors (i.e., HMD controller battery issue, tracking loss issue after a rest phase). However, several features need to be improved in the future. For instance, we could add functions that manage the trial process, such as inter-trial interval (ITI) and stimuli presentation time.

Our library could contribute to inviting researchers from various backgrounds to virtual and augmented reality domain research.  By relieving their burden on programming in Unity and helping them focus on experiment design, we expect the research that utilizes VR/AR environments will be expanded and new perspectives and improvements in the academic field could be gained.

\bibliographystyle{unsrtnat}
\bibliography{exptrialmng}  






\end{document}